# Quantum Sensing of Spin Fluctuations of Magnetic Insulator Films with Perpendicular Anisotropy

Eric Lee-Wong[1,2], Jinjun Ding[3], Xiaoche Wang[1], Chuanpu Liu[3], Nathan J. McLaughlin[1], Hailong Wang[4], Mingzhong Wu[3], Chunhui Rita Du[1,4]

[1]Department of Physics, University of California, San Diego, La Jolla, California 92093
[2]Department of NanoEngineering, University of California, San Diego, La Jolla, California 92093
[3]Department of Physics, Colorado State University, Fort Collins, Colorado 80523
[4]Center for Memory and Recording Research, University of California, San Diego, La Jolla, California 92093

Nitrogen vacancy (NV) centers, optically active atomic defects in diamond, have been widely applied to emerging quantum sensing, imaging, and network efforts, showing unprecedented field sensitivity and nanoscale spatial resolution. Many of these advantages derive from their excellent quantum-coherence, controllable entanglement, and high fidelity of operations, enabling opportunities to outperform the classical counterpart. Exploiting this cutting-edge quantum metrology, we report noninvasive measurement of intrinsic spin fluctuations of magnetic insulator thin films with a spontaneous out-of-plane magnetization. The measured field dependence of NV relaxation rates is well correlated to the variation of magnon density and band structure of the magnetic samples, which are challenging to access by the conventional magnetometry methods. Our results highlight the significant opportunities offered by NV centers in diagnosing the noise environment of functional magnetic elements, providing valuable information to design next-generation, high-density, and scalable spintronic devices.

Magnetic thin films with perpendicular anisotropy are widely used for information processing and data storage applications.[1,2] The spontaneous out-of-plane magnetization enables significant improvement of density, speed, and reliability in a variety of spin-based electronic devices, such as magnetic tunnel junctions,[3] spin-torque nano-oscillators,[4] and racetrack memory.[5–7] A range of material candidates including rare-earth/transition-metal alloys,[8,9] L1$_0$-ordered (Co, Fe)–Pt alloys,[10] and Co/(Pd, Pt) multilayers[11,12] have been extensively explored towards this end. More recently, the material landscape has extended to insulators, e. g. Y$_3$Fe$_5$O$_{12}$,[13–15] Tm$_3$Fe$_5$O$_{12}$,[16–18] and BaFe$_{12}$O$_{19}$ thin films,[19] whose perpendicular magnetic anisotropy (PMA) results from epitaxial strain or intrinsic magneto-crystalline anisotropy. Remarkably, the Gilbert damping of magnetic insulators is orders of magnitude smaller in comparison with their metallic counterpart, providing an excellent material platform for developing energy-efficient magnetic switching,[16,17] domain wall motions,[6,7] and long-range spin information transmission.[20,21]

Despite these potential benefits, a direct measurement of the intrinsic spin fluctuations of insulating magnetic films with PMA is missing. In modern spintronic technologies, the magnitude and spatial profile of magnetic noise generated by a nanomagnet play an important role in determining the density and scalability of the miniaturized electronic devices.[22,23] The magnetic noise is associated with the internal spin/magnon fluctuations, which are related to the underlying magnetic properties, such as magnon-phonon interactions,[24,25] magnon scattering,[26–29] and magnon hydrodynamics.[30] Therefore, detailed information on the key material properties and device parameters can be extracted non-invasively by probing the intrinsic magnetic noise spectrum. Conventional magnetometry methods such as vibrating sample magnetometer,[31] SQUID,[32] and ferromagnetic resonance (FMR)[33] are mainly designed to characterize the static and/or coherent dynamic magnetic properties, and are unable to access the low magnitude non-coherent magnetic fluctuations. To address this challenge, here, we introduce nitrogen vacancy (NV) centers,[34] optically-active atomic defects in diamond, to locally probe the magnetic noise generated by nanometer-thick, perpendicularly magnetized Y$_3$Fe$_5$O$_{12}$ (YIG) thin films. The observed field-dependent NV relaxation rates are well explained by the variation of the magnon density and band structure of the samples. The sensitivity length scale of the presented quantum sensing platform is mainly determined by the NV-to-sample distance, which can ultimately approach the tens-of-nanometer regime.[35,36] Our results highlight the significant opportunities offered by NV centers in diagnosing the intrinsic spin fluctuations in a broad range of functional magnetic systems.

We start by discussing the structural and magnetic properties of the epitaxial YIG thin films with PMA. The YIG films are grown on single-crystal (111) Gd$_3$(Sc$_2$Ga$_3$)O$_{12}$ (GSGG) substrates by radio-frequency sputtering. Detailed information of the growth conditions has been reported in the previous work.[14] Figure 1(a) shows the representative $\theta - 2\theta$ x-ray diffraction scans of a series of YIG films with different thicknesses and a GSGG substrate. The two main peaks around 50.4 degree correspond to the (444) peaks of the GSGG substrate. The appearance of the two (444) peaks result from the coexistence of the $K_{\alpha 1}$ and $K_{\alpha 2}$ components of the x ray. The black arrows indicate the positions of the (444) peak of the YIG films, which clearly deviate from the bulk value as marked by the vertical dashed line. The lattice constant of GSGG is 12.554 Å, which is slightly larger than the bulk value of YIG which is 12.376 Å. The tensile strain provided by the GSGG substrate induces a substantial out-of-plane magnetocrystalline anisotropy in the YIG thin film, which rotates the magnetic easy axis from in-plane (IP) to out-of-plane (OOP) direction. Figure 1(b) shows the field dependent magnetization of 8-nm and 12-nm-thick YIG thin films measured with an external magnetic field $H$ applied in the OOP direction. The paramagnetic background of the GSGG substrate has been subtracted. The nearly square hysteresis loop with a high remnant-to-saturation magnetization ratio demonstrates the PMA of the YIG thin films. An atomic force microscopy image in Fig. 1(c) shows a smooth surface of the YIG (8 nm)/GSGG sample with a roughness of 0.12 nm.

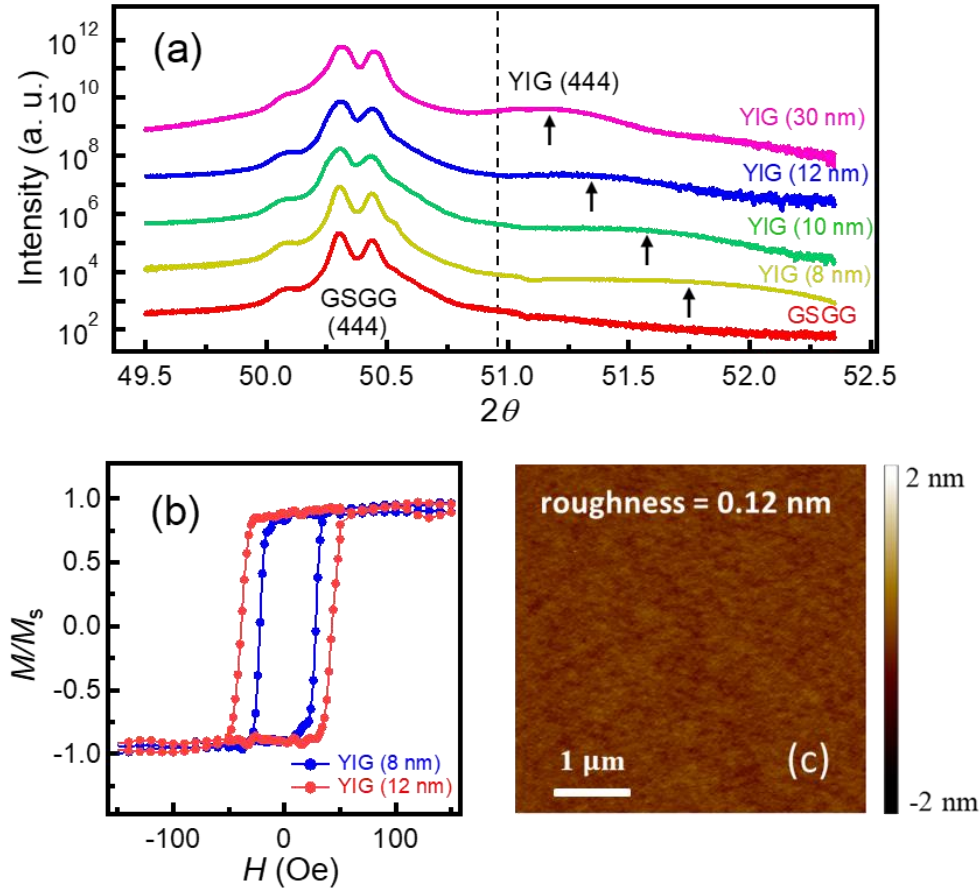

**Figure 1**. (a) X-ray diffraction (XRD) $\theta$-$2\theta$ scan of 8-nm, 10-nm, 12-nm, and 30-nm-thick YIG films grown on GSGG (111) substrates. The XRD data of the substrate is also shown for comparison. The curves are offset for clarity. (b) Out-of-plane magnetic hysteresis loops for YIG(8 nm)/GSGG and YIG(12 nm)/GSGG samples. (c) Atomic force microscopy image of an 8-nm-thick YIG film grown on GSGG (111) over an area of 5 μm × 5 μm, showing a surface roughness of 0.12 nm.

The strong PMA will change the effective magnetization $4\pi M_{\text{eff}}$ of the YIG thin films as follows: $4\pi M_{\text{eff}} = 4\pi M_s - H_\perp$,[13] where $4\pi M_s$ is the saturation magnetization and $H_\perp$ is PMA. To quantitatively characterize this effect, we performed angle and frequency dependent FMR measurements to extract $4\pi M_{\text{eff}}$ of the prepared YIG/GSGG samples. Figure 2(a) shows two representative FMR spectra of the 8-nm-thick PMA YIG film measured at a microwave frequency $f = 10$ GHz and a polar angle $\theta_H = 0$ and 90 degrees. Here, $\theta_H$ is defined to be the angle between the normal of the film plane and the external magnetic field. A significantly lower FMR resonant field $H_{res}$ measured when $\theta_H = 0$ degree confirms the OOP orientated magnetic easy axis. Figure 2(b) shows the angular dependence of $H_{res}$ for 8 nm and 12 nm PMA YIG thin films measured at $f = 10$ GHz. By fitting to Fig. 2(b), the effective magnetization $4\pi M_{eff}$ of YIG (8 nm)/GSGG and YIG (12 nm)/GSGG is extracted to be $-456 \pm 7$ Oe and $-1489 \pm 10$ Oe, respectively (see supplementary information for details). The negative sign results from the strong PMA, which overcomes the intrinsic YIG magnetization and dictates the magnetic easy axis. We also measured the frequency dependent $H_{res}$ of the PMA YIG films in both OOP and IP field geometries, as shown in Fig. 2(c). The measurement results can be well fitted to the Kittel equations: $f = \frac{\gamma}{2\pi}(H - 4\pi M_{\text{eff}})$ ($\theta_H = 0$ degree) and $f = \frac{\gamma}{2\pi}\sqrt{H(H + 4\pi M_{\text{eff}})}$ ($\theta_H = 90$ degree), where $\gamma$ is the gyromagnetic ratio of the magnetic sample. The obtained $4\pi M_{\text{eff}}$ is in agreement with the angle dependent FMR measurement results. In order to determine the Gilbert damping of the YIG thin films, we measured the frequency dependence of the FMR linewidth $\Delta H_{\text{FMR}}$ as shown in Fig. 2(d). In all cases, $\Delta H_{\text{FMR}}$ increases linearly with $f$. The Gilbert damping constant $\alpha$ and the inhomogeneous linewidth broadening $\Delta H_{\text{inh}}$ can be obtained using equation:[37,38]

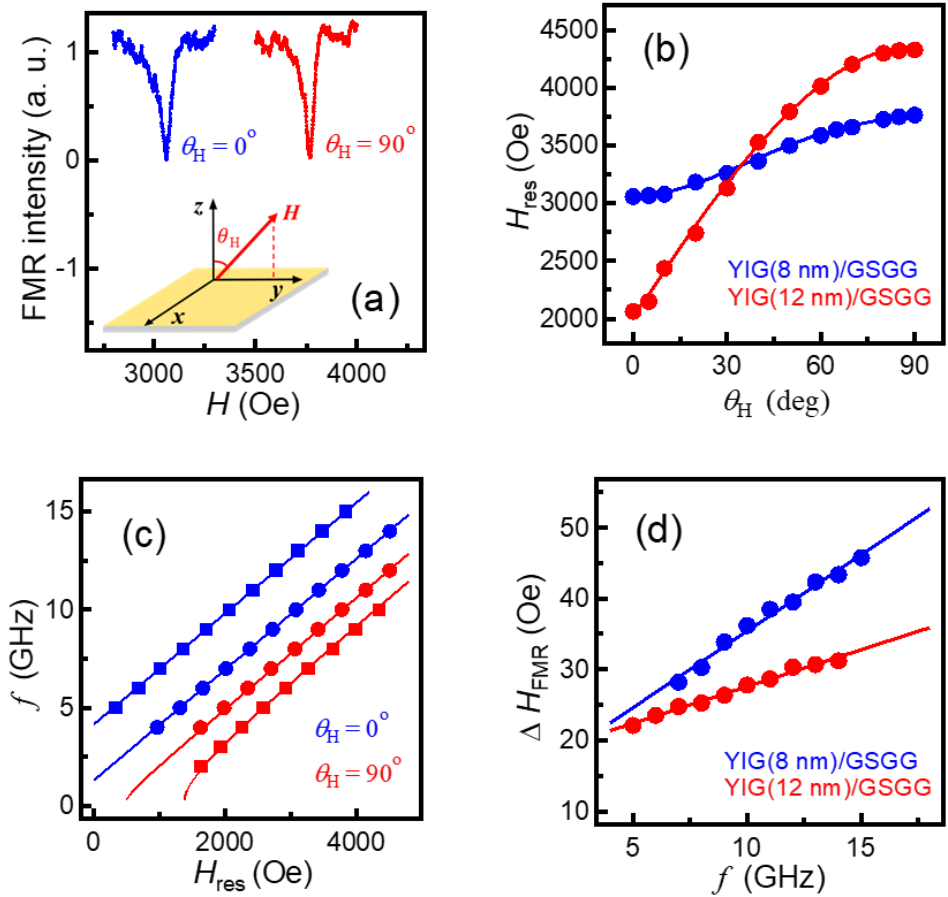

**Figure 2**. (a) Two representative FMR spectra (in arbitrary unit) at a frequency $f = 10$ GHz for the 8-nm-thick YIG film grown GSGG (111). The red and blue curves correspond to in-plane and out-of-plane magnetic field geometries, respectively. (b) Angular dependence of the obtained FMR resonant field $H_{res}$ for YIG(8 nm)/GSGG (blue points) and YIG(12 nm)/GSGG (red points). The curves are fittings to extract the effective magnetization. (c) Frequency dependence of $H_{res}$ for YIG(8 nm)/GSGG (dots) and YIG(12 nm)/GSGG (squares) measured in both in-plane (red color) and out-of-plane (blue color) field geometries. The red and blue curves are fittings to Kittel equations. (d) Frequency dependence of FMR linewidth $\Delta H_{FMR}$ for 8-nm (blue points) and 12-nm (red points) PMA YIG films. The red and blue curves are linear fittings to extract Gilbert damping and inhomogeneous contribution to $\Delta H_{FMR}$.

$$\Delta H_{\text{FMR}} = \Delta H_{\text{inh}} + \frac{4\pi \alpha f}{\sqrt{3}\gamma} \tag{1}$$

Table I summaries the obtained magnetic properties of the 8-nm and 12-nm-thick YIG films grown on GSGG (111) substrates. Our NV quantum sensing measurements are mainly focused on these two samples showing a large variation of the effective magnetization.

Next, we employ NV centers to perform local sensing of the intrinsic spin fluctuations of the YIG (8 nm)/GSGG sample. An NV center is formed by a nitrogen atom adjacent to a carbon atom vacancy in one of the nearest neighboring sites of a diamond crystal lattice.[34] The negatively charged NV state has an $S = 1$ electron spin and serves as a three-level qubit system. Due to their excellent quantum coherence and single-spin sensitivity, NV centers have been successfully applied to quantum sensing and imaging research, showing unprecedented field sensitivity and nanoscale spatial solution.[39,40] Figure 3(a) illustrates the material and device structure for our NV-based quantum sensing measurements. A patterned diamond nanobeam[27,41,42] containing single NV spins is mechanically transferred onto the surface of a YIG/GSGG sample. The diamond nanobeam has the shape of an equilateral triangular prism with dimensions of 500 nm × 500 nm × 10 μm. It is in van der Waals contact with the sample surface, ensuring nanoscale proximity between NV centers and the YIG film. A photoluminescence (PL) image [Fig. 3(b)] shows individual NV centers positioned on top of the sample surface, demonstrating single-spin addressability of our confocal measurement system. An external magnetic field is applied and aligned to the NV-axis with an angle $\theta_{\text{NV}}$ relative to the surface normal.

At thermal equilibrium, transverse spin fluctuations of YIG magnons generate dipolar stray fields at the NV site. When the frequency of YIG thermal magnons equals the NV electron spin resonance (ESR) frequency, the magnetic noise will induce NV spin transitions from the $m_s = 0$ to the $m_s = \pm 1$ states, leading to an enhanced NV relaxation rate (see supplementary information for details):[27,43]

$$\Gamma_{\pm} = \frac{k_B T}{f_{\pm} h} \int D(f_{\pm}, \boldsymbol{k}) f(\boldsymbol{k}, d) \, d\boldsymbol{k} \tag{2}$$

where $f_{\pm}$ and $\Gamma_{\pm}$ are the NV ESR frequencies and NV relaxation rates of the $m_s = 0 \leftrightarrow \pm 1$ transitions respectively, $T$ is the temperature, $k_B$ is the Boltzmann constant, $h$ is the Planck constant, $D(f_{\pm}, \boldsymbol{k})$ is magnon spectral density, $\boldsymbol{k}$ is the magnon wave vector, and $f(\boldsymbol{k}, d)$ is transfer function describing the magnon-generated stray fields at the NV site. Due to the "filtering" effect of $f(\boldsymbol{k}, d)$, an NV center is mostly sensitive to magnetic noise with a wavevector $k \sim 1/d$, where $d$ characterizes the distance between the surface of the magnetic sample and the NV spin (see supplementary information for details). The magnetic fields with a smaller wavevector are algebraically suppressed and vanish at $k = 0$, which is expected because a uniform magnetization generates zero stray field. When $k$ is larger than $1/d$, the dipolar fields are exponentially suppressed with an increasing $k$ due to the self-averaging of short-wavelength fluctuations.

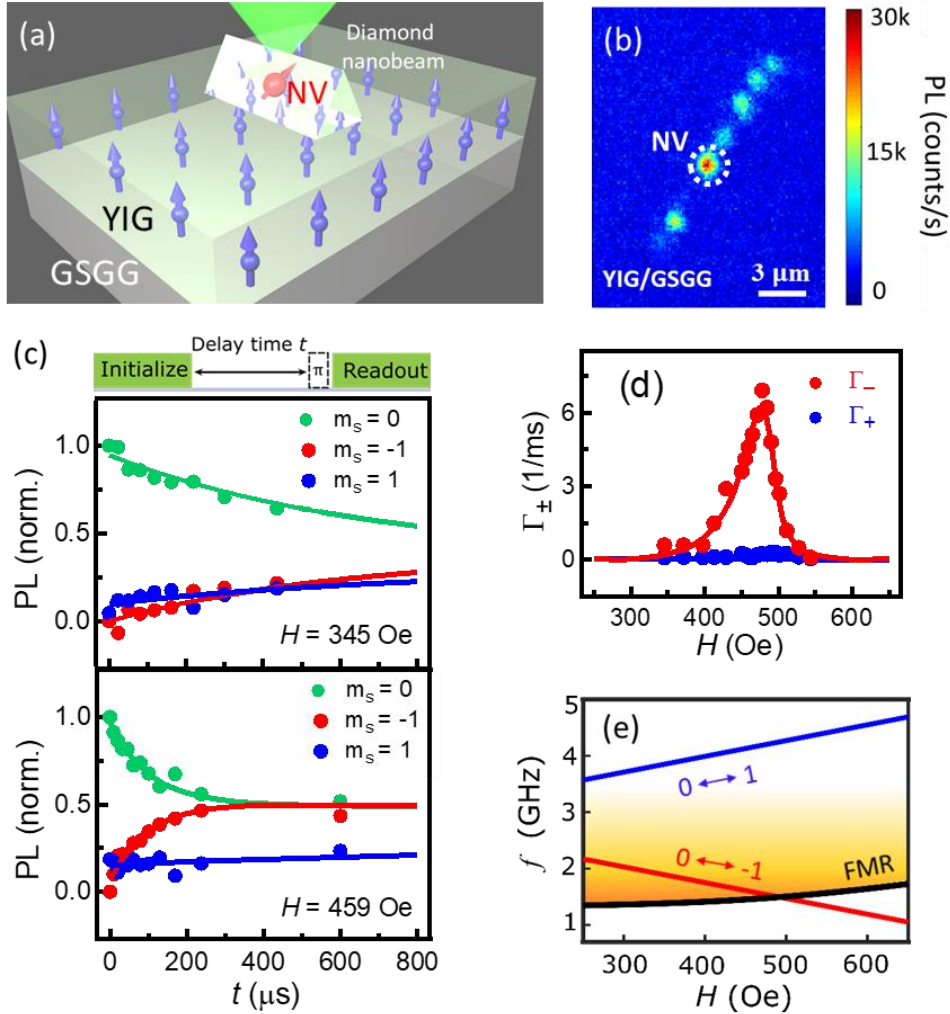

**Figure 3**. (a) Schematic of a single NV spin contained in a patterned diamond nanobeam locally probing the spin fluctuations of a PMA YIG thin film grown on a GSGG substrate. (b) A photoluminescence image showing individually addressable NV spins positioned on top of the sample surface. (c) Top panel: optical and microwave sequence for NV relaxometry measurements. Bottom panel: two sets of NV relaxation data measured with $H = 345$ Oe and 459 Oe for an 8-nm-thick YIG film grown GSGG. (d) NV spin relaxation rate $\Gamma_+$ (blue points) and $\Gamma_-$ (red points) measured as a function of $H$. The solid lines are fitting to Eq. (2), which gives the NV-to-sample distance of $239 \pm 11$ nm. (e) Sketch of the magnon density of the YIG (8 nm)/GSGG sample and the NV ESR frequencies $f_\pm$ as a function of $H$.

Experimentally, we employed NV relaxometry measurements[27,28,44] to detect the spin fluctuations of the PMA YIG samples. The top panel of Fig. 3(c) shows the optical and microwave measurement sequence. A green laser pulse is first applied to initialize the NV spin to the $m_s = 0$ state. The spin noise generated by YIG magnons at frequencies $f_\pm$ induces NV spin transitions from the $m_s = 0$ to the $m_s = \pm 1$ states. After a delay time $t$, we measure the occupation probabilities of the NV spin at the $m_s = 0$ and the $m_s = \pm 1$ states by applying a microwave $\pi$ pulse on the corresponding ESR frequencies and measuring the spin-dependent PL during the green-laser readout pulse. By measuring the integrated PL intensity as a function of the delay time $t$, NV relaxation rates can be quantitatively obtained by fitting to a three-level model.[27,43] The bottom panel of Fig. 3(c) shows two sets of NV spin relaxation data measured with $H$ = 345 Oe and 459 Oe. The measured PL intensity corresponding to the $m_s = 0$ ($\pm 1$) state decreases (increases) as a function of $t$, demonstrating the magnetic-noise-induced relaxation of the NV spin to a mixture of the $m_s = 0$ and the $m_s = \pm 1$ states.

Figure 3(d) shows the obtained NV spin relaxation rates $\Gamma_\pm$ as a function of $H$. The results can be well fitted by Eq. (2), by which the NV-to-sample distance $d$ is obtained to be $239 \pm 11$ nm. Notably, $\Gamma_-$ corresponding to the $m_s = 0 \leftrightarrow -1$ transition exhibits a significant variation with $H$ and shows a maximal value around $H$ = 477 Oe. In contrast, the magnitude of $\Gamma_+$ corresponding to the $m_s = 0 \leftrightarrow +1$ transition is orders of magnitude smaller in comparison with $\Gamma_-$. The variation of the measured NV relaxation rates is correlated to the field-dependent magnon density of the sample and the NV ESR frequencies $f_\pm$. Figure 3(e) plots the magnon density of the 8-nm-thick YIG thin film and $f_\pm$ as a function of $H$. The magnon density falls off as 1/energy (1/frequency) as indicated by the fading color. The minimal magnon band is determined by the FMR frequency $f_{\text{FMR}}$, which is calculated by the magnon dispersion relationship with $\theta_{\text{NV}}$ = 72 degree (see supplementary information for details). Due to the Zeeman splitting, the NV ESR frequencies $f_\pm$ vary with the external magnetic field as follows: $f_\pm = 2.87 \pm \tilde{\gamma} H/2\pi$, where $\tilde{\gamma}$ is the gyromagnetic ratio of the NV spin. As $H$ increases, the lower branch of the NV ESR frequency $f_-$ decreases and probes the YIG thermal magnons with a higher density, leading to a significant enhancement of $\Gamma_-$ as shown in Fig. 3(d). The maximal $\Gamma_-$ emerges at $H \sim 477$ Oe, where $f_-$ corresponds to the peak regime of the "filter function" (see supplementary information for details). When $H > 492$ Oe, $f_-$ lies below the minimum of YIG magnon band. The measured $\Gamma_-$ rapidly decays, indicating effective suppression of spin noise generated at frequency $f_-$. For the upper branch of the NV ESR frequency $f_+$, it probes YIG thermal magnons with higher frequencies, lower magnon densities and wavevectors falling outside the NV sensitivity regime, therefore, $\Gamma_+$ is orders of magnitude smaller than $\Gamma_-$ and does not exhibit a significant variation with $H$.

The magnetic properties of perpendicularly magnetized YIG thin films are usually sensitive to epitaxial strain, thickness, and detailed growth parameters as shown in Table I. To illustrate the versatility of the NV quantum sensing platform, we also apply NV relaxometry measurements to the YIG (12 nm)/GSGG sample, whose zero-field magnon band gap is larger than the corresponding NV ESR frequency. Figures 4(a) and 4(b) show two sets of NV spin relaxation data measured with $H$ = 548 Oe and 736 Oe, respectively. The extracted field-dependent NV relaxation rates $\Gamma_\pm$ are plotted in Fig. 4(c). Due to the significantly enhanced PMA ($4\pi M_{\text{eff}} = -1489 \pm 10$ Oe), the minimal magnon energy $f_{\text{FMR}}$ of the 12-nm-thick PMA YIG film is 3.9 GHz, which is above $f_\pm = 2.87$ GHz when $H$ = 0. As $H$ increases, $f_-$ always lies below the magnon band minimum, thus, giving rise to a negligible NV relaxation rate $\Gamma_-$ as shown in Fig. 4(c). In contrast, the upper branch of the NV ESR frequency $f_+$ increases faster than $f_{\text{FMR}}$ and crosses with the magnon band minimum at $H$ = 727 Oe [Fig. 4(d)], accompanied with a dramatic enhancement of $\Gamma_+$. Note that the maximal NV relaxation rate is about one order of magnitude larger than the value observed in the case of the 8-nm-thick YIG sample, which mainly results from a closer NV-to-sample distance ($d$ = 114 ± 10 nm). As $f_+$ further increases, the NV relaxation rate $\Gamma_+$ starts to decay due to a reduced magnon density and the "filtering" effect of the wavevector.[27,43]

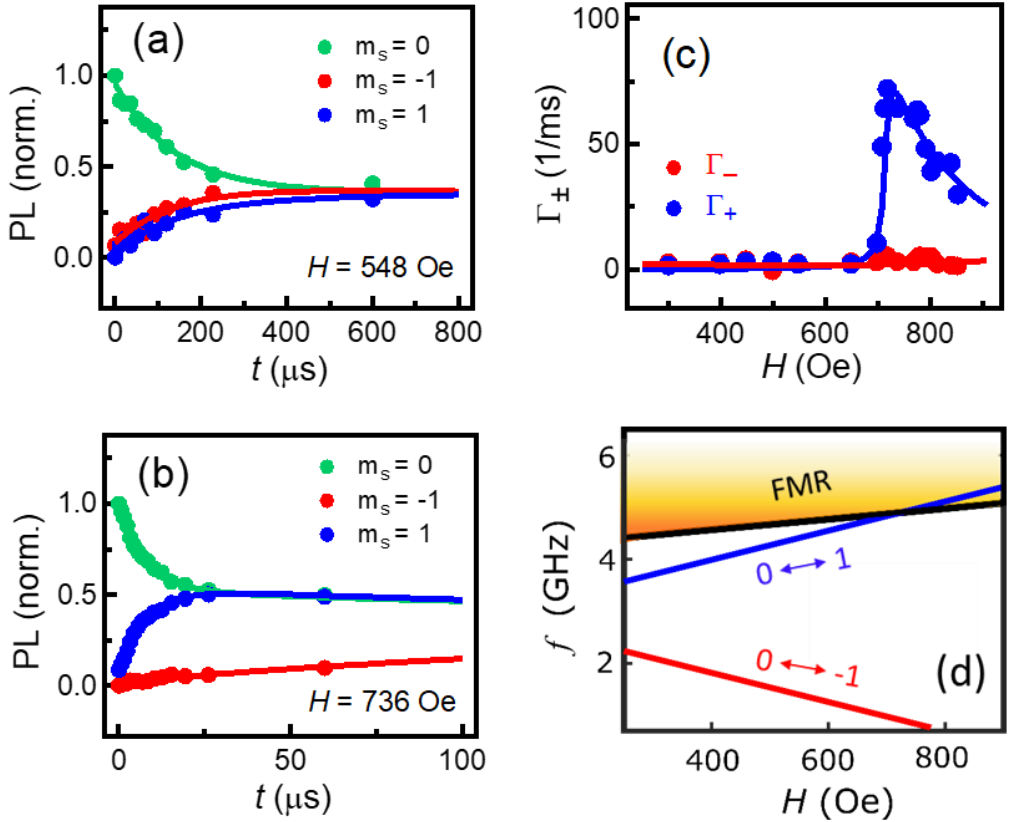

**Figure 4**. (a, b) Two sets of NV relaxation data measured with $H$ = 548 Oe and 736 Oe for a 12-nm-thick YIG film grown a GSGG substrate. (c) NV spin relaxation rate $\Gamma_+$ (blue points) and $\Gamma_-$ (red points) measured as a function of $H$. The solid lines are fitting to Eq. (2), which gives the NV-to-sample distance of 114 ± 10 nm. (d) Sketch of the magnon density of the YIG(12 nm)/GSGG sample and the NV ESR frequencies $f_\pm$ as a function of $H$.

In summary, we have demonstrated NV centers as a local probe of the intrinsic spin fluctuations of nanometer-thick PMA YIG thin films. The observed field-dependent NV relaxation rates are well explained by the variation of the magnon density, magnon band gaps of the magnetic samples, and the NV transfer function. In contrast to the conventional magnetic resonance techniques that mainly probe the coherent spin wave modes, we highlight that NV relaxometry provides a unique way to access noncoherent magnon thermal fluctuations in a noninvasive fashion. We expect that three-dimensional mapping of magnetic noise of functional spintronic devices can be ultimately achieved by employing scanning NV microscopy.[35,36] The demonstrated coupling between perpendicularly magnetized ferromagnets and NV centers may also find applications in building solid-state-based hybrid quantum architectures for next-generation spintronic technologies.[45–48,49]

**Acknowledgements**. The work at UCSD was supported by the U. S. National Science Foundation under award ECCS-2029558 and the Air Force Office of Scientific Research under award FA9550-20-1-0319. The work at CSU was supported by the U.S. National Science Foundation under Grants No. EFMA-1641989 and No. ECCS-1915849.

# References

1. S. Ikeda *et al.*, A perpendicular-anisotropy CoFeB-MgO magnetic tunnel junction, *Nat. Mater.* **9**, 721 (2010).
2. B. Tudu and A. Tiwari, Recent developments in perpendicular magnetic anisotropy thin films for data storage applications, *Vacuum* **146**, 329 (2017).
3. J. S. Moodera, L. R. Kinder, T. M. Wong, and P. Meservey, Large magnetoresistance at room temperature in ferromagnetic thin film tunnel junctions, *Phys. Rev. Lett.* **74**, 3273 (1995).
4. T. Chen *et al.*, Spin-torque and spin-Hall nano-oscillators, *Proceedings of the IEEE* **104**, 1919 (2016).
5. S. S. P. Parkin, M. Hayashi, and L. Thomas, Magnetic domain-wall racetrack memory, *Science* **320**, 190 (2008).
6. S. Emori, U. Bauer, S. M. Ahn, E. Martinez, and G. S. D. Beach, Current-driven dynamics of chiral ferromagnetic domain walls, *Nat. Mater.* **12**, 611 (2013).
7. S. Vélez *et al.*, High-speed domain wall racetracks in a magnetic insulator, *Nat. Commun.* **10**, 4750 (2019).
8. N. Nishimura *et al.*, Magnetic tunnel junction device with perpendicular magnetization films for high-density magnetic random access memory, *J. Appl. Phys.* **91**, 5246 (2002).
9. J. Finley and L. Liu, Spin-orbit-torque efficiency in compensated ferrimagnetic Cobalt-Terbium alloys, *Phys. Rev. Appl.* **6**, 054001 (2016).
10. G. Kim, Y. Sakuraba, M. Oogane, Y. Ando, and T. Miyazaki, Tunneling magnetoresistance of magnetic tunnel junctions using perpendicular magnetization $L1_0$-CoPt electrodes, *Appl. Phys. Lett.* **92**, 172502 (2008).
11. S. Mangin *et al.*, Current-induced magnetization reversal in nanopillars with perpendicular anisotropy, *Nat. Mater.* **5**, 210 (2006).
12. B. Carvello *et al.*, Sizable room-temperature magnetoresistance in cobalt based magnetic tunnel junctions with out-of-plane anisotropy, *Appl. Phys. Lett.* **92**, 102508 (2008).
13. H. L. Wang, C. H. Du, P. C. Hammel, and F. Yang, Strain-tunable magnetocrystalline anisotropy in epitaxial $Y_3Fe_5O_{12}$ thin films, *Phys. Rev. B.* **89**, 134404 (2014).
14. J. Ding *et al.*, Nanometer-thick Yttrium iron garnet films with perpendicular anisotropy and low damping, *Phys. Rev. Appl.* **14**, 014017 (2020).
15. L. Soumah *et al.*, Ultra-low damping insulating magnetic thin films get perpendicular, *Nat. Commun.* **9**, 3355 (2018).
16. Q. Shao *et al.*, Role of dimensional crossover on spin-orbit torque efficiency in magnetic insulator thin films, *Nat. Commun.* **9**, 3612 (2018).
17. C. O. Avci *et al.*, Current-induced switching in a magnetic insulator, *Nat. Mater.* **16**, 309 (2017).
18. A. J. Lee *et al.*, Probing the source of the interfacial Dzyaloshinskii-Moriya interaction responsible for the topological Hall effect in metal/$Tm_3Fe_5O_{12}$ systems, *Phys. Rev. Lett.* **124**, 107201 (2020).
19. P. Li *et al.*, Magnetization switching using topological surface states, *Sci. Adv.* **5**, eaaw3415 (2019).
20. R. Lebrun *et al.*, Tunable long-distance spin transport in a crystalline antiferromagnetic iron oxide, *Nature* **561**, 222 (2018).
21. L. J. Cornelissen, J. Liu, R. A. Duine, J. Ben Youssef, and B. J. Van Wees, Long distance transport of magnon spin information in a magnetic insulator at room temperature, *Nat. Phys.* **11**, 1022 (2015).

22. S. Chatterjee, J. F. Rodriguez-Nieva, and E. Demler, Diagnosing phases of magnetic insulators via noise magnetometry with spin qubits, *Phys. Rev. B* **99**, 104425 (2019).
23. P. Upadhyaya *et al.*, Thermal stability characterization of magnetic tunnel junctions using hard-axis magnetoresistance measurements, *J. Appl. Phys.* **109**, 07C708 (2011).
24. J. Holanda, D. S. Maior, A. Azevedo, and S. M. Rezende, Detecting the phonon spin in magnon-phonon conversion experiments, *Nat. Phys.* **14**, 500 (2018).
25. K. An *et al.*, Magnons and phonons optically driven out of local equilibrium in a magnetic insulator, *Phys. Rev. Lett.* **117**, 107202 (2016).
26. S. O. Demokritov *et al.*, Bose-Einstein condensation of quasi-equilibrium magnons at room temperature under pumping, *Nature* **443**, 430 (2006).
27. C. H. Du *et al.*, Control and local measurement of the spin chemical potential in a magnetic insulator, *Science* **357**, 195 (2017).
28. M. R. Page *et al.*, Optically detected ferromagnetic resonance in metallic ferromagnets via nitrogen vacancy centers in diamond, *J. Appl. Phys.* **126**, 124902 (2019).
29. B. Flebus and Y. Tserkovnyak, Quantum-impurity relaxometry of magnetization dynamics, *Phys. Rev. Lett.* **121**, 187204 (2018).
30. J. F. Rodriguez-Nieva, D. Podolsky, and E. Demler, Hydrodynamic sound modes and viscous damping in a magnon fluid, http://arxiv.org/abs/1810.12333 (2020).
31. S. Foner, Versatile and sensitive vibrating-sample magnetometer, *Rev. Sci. Instrum.* **30**, 548 (1959).
32. M. Buchner, K. Höfler, B. Henne, V. Ney, and A. Ney, Tutorial: Basic principles, limits of detection, and pitfalls of highly sensitive SQUID magnetometry for nanomagnetism and spintronics, *J. Appl. Phys.* **124**, 161101 (2018).
33. C. Kittel, On the theory of ferromagnetic resonance absorption, *Phys. Rev.* **73**, 155 (1948).
34. L. Rondin *et al.*, Magnetometry with nitrogen-vacancy defects in diamond, *Reports Prog. Phys.* **77**, 056503 (2014).
35. M. Pelliccione *et al.*, Scanned probe imaging of nanoscale magnetism at cryogenic temperatures with a single-spin quantum sensor, *Nat. Nanotechnol.* **11**, 700 (2016).
36. L. Thiel *et al.*, Probing magnetism in 2D materials at the nanoscale with single-spin microscopy, *Science* **364**, 973 (2019).
37. R. D. McMichael, D. J. Twisselmann, and A. Kunz, Localized ferromagnetic resonance in inhomogeneous thin films, *Phys. Rev. Lett.* **90**, 4 (2003).
38. J. M. Shaw, H. T. Nembach, and T. J. Silva, Determination of spin pumping as a source of linewidth in sputtered $Co_{90}Fe_{10}$/Pd multilayers by use of broadband ferromagnetic resonance spectroscopy, *Phys. Rev. B.* **85**, 054412 (2012).
39. P. Neumann *et al.*, High-precision nanoscale temperature sensing using single defects in diamond, *Nano Lett.* **13**, 2738 (2013).
40. G. Balasubramanian *et al.*, Nanoscale imaging magnetometry with diamond spins under ambient conditions, *Nature* **455**, 648 (2008).
41. M. J. Burek *et al.*, Free-standing mechanical and photonic nanostructures in single-crystal diamond, *Nano Lett.* **12**, 6084 (2012).
42. E. Lee-Wong, R. L. Xue, F. Y. Ye, A. Kreisel, T. van der Sar, A. Yacoby, and C. H. R. Du, Nanoscale detection of magnon excitations with variable wavevectors through a quantum spin sensor, *Nano Lett.* **20**, 3284 (2020).
43. T. van der Sar, F. Casola, R. Walsworth, and Amir Yacoby, Nanometre-scale probing of spin waves using single-electron spins, *Nat. Commun.* **6**, 7886 (2015).

44. J. P. Tetienne *et al.*, Spin relaxometry of single nitrogen-vacancy defects in diamond nanocrystals for magnetic noise sensing, *Phys. Rev. B.* **87**, 235436 (2013).
45. X. Wang *et al.*, Electrical control of coherent spin rotation of a single-spin qubit, http://arxiv.org/abs/2007.07543 (2020).
46. P. Andrich *et al.*, Long-range spin wave mediated control of defect qubits in nanodiamonds, *npj Quantum Inf.* **3**, 28 (2017).
47. E. R. MacQuarrie, T. A. Gosavi, N. R. Jungwirth, S. A. Bhave, and G. D. Fuchs, Mechanical spin control of nitrogen-vacancy centers in diamond, *Phys. Rev. Lett.* **111**, 227602 (2013).
48. D. Labanowski *et al.*, Voltage-driven, local, and efficient excitation of nitrogen-vacancy centers in diamond, *Sci. Adv.* **4**, eaat6574 (2018).
49. D. Kikuchi *et al.*, Long-distance excitation of nitrogen-vacancy centers in diamond via surface spin waves, *Appl. Phys. Express* **10**, 103004 (2017).

| Thickness (nm) | $4\pi M_{eff}$ (Oe) | Gilbert damping $\alpha$ ($\times 10^{-3}$) | Inhomogeneous linewidth broadening $\Delta H_{inh}$ (Oe) |
|---|---|---|---|
| 8 | -456 ± 7 | 5.2 ± 0.2 | 13.8 ± 1.1 |
| 12 | -1489 ± 10 | 2.5 ± 0.1 | 17.2 ± 0.4 |

**Table I.** Magnetic properties of YIG thin films grown GSGG substrates.